\documentclass[preprint,12pt,authoryear]{elsarticle}

\usepackage{amssymb}
\usepackage{color}
\usepackage{bm}        
\usepackage{latexsym,amsmath}
\usepackage{dcolumn}   
\usepackage{natbib}
\usepackage{graphicx}  
\usepackage{float}
\usepackage{bm}        
\usepackage{isomath}
\usepackage{chngcntr} 
\usepackage{secdot} 

\journal{Journal of the Mechanics and Physics of Solids}

\begin{document}

\begin{frontmatter}



\title{Nonlinear Wave Propagation in 3D-Printed Graded Lattices of Hollow Elliptical Cylinders}


\author[1]{Hyunryung Kim}
\author[2]{Eunho Kim}
\author[1]{Jinkyu Yang \corref{3}}

\address[1]{Aeronautics and Astronautics, University of Washington, Seattle, WA, USA, 98195-2400}
\address[2]{Division of Mechanical System Engineering \& Automotive Hi-Technology Research Center, Chonbuk National University, 567 Baekje-daero, Deokjin-gu, Jeonju-si, Jeollabuk-do 54896, Republic of Korea}
\cortext[3]{Author to which all correspondence should be addressed: jkyang@aa.washington.edu}

\begin{abstract}
We propose a 3D-printed graded lattice made of hollow elliptical cylinders (HECs) as a new way to design impact mitigation systems.
We observe asymmetric dynamics in the graded HEC chains with increasing and decreasing stiffness. 
Specifically, the increasing stiffness chain shows an acceleration of the propagating waves, while the decreasing stiffness chain shows the opposite. 
From the standpoint of impact mitigation, the decreasing stiffness chain combined with the strain-softening behavior of HECs results in an order-of-magnitude improvement in force attenuation compared to the increasing stiffness chain.
We extend this finding to the graded 2D arrays and demonstrate a similar trend of wave transmission efficiency contrast between the increasing and decreasing stiffness lattices. The 3D-printed HEC lattices shown in this study can lead to the development of a new type of impact mitigating and shock absorbing structures.   
\end{abstract}

\begin{keyword}
Nonlinear \sep 3D printing \sep Stress wave \sep Tunability \sep Impact mitigation
\end{keyword}

\end{frontmatter}

\section{Introduction}
Tapered or graded granular chains have been of great interest to researchers for their exceptional impact absorbing ability \citep{Sen2008,Rosas2018}. First proposed by \cite{Sen2001}, long tapered granular chains (total number of granular particles, or chain length, $N = 100$) have been shown to reduce their leading pulse's kinetic energy up to 90\% compared to monodispersed chains in numerical analysis. Later on, analytic models have been developed \citep{Wu2002,Doney2005,Harbola2009Tap,Machado2013} to describe the physics of the remarkable energy attenuation. Numerical \citep{Sokolow2005} and experimental demonstration \citep{Nakagawa2003,Melo2006} have also been reported to claim the feasibility in shorter chains ($N \approx 20$).

\cite{Doney2006} introduced a superb shock absorbing system, so-called decorated tapered chain (DTC), which mitigates the impact energy significantly in a very short chain length. Supporting experimental results \citep{Doney2009} and analytic interpretations \citep{Harbola2009Dec} have been reported. The challenge with the DTC is that DTC itself is not feasible to be assembled in higher dimensions due to their geometrical constraints.
\cite{Machado2014} found an alternative to DTC by surrounding the regular tapered chain with smaller DTCs. Impulse wave propagation is more attenuated in this quasi-one-dimensional (1D) tapered chain than the regular tapered chain. Ideas of stacking tapered granular chains into two-dimensional (2D) \citep{Tiwari2016} and three-dimensional (3D) \citep{Sen2017} space have been proposed recently. However, these studies on tapered granular structures in higher dimensions are limited to numerical investigation, and their experimental demonstration is yet elusive, mainly due to the challenges in their assembly.

3D-printing techniques have been recently gaining their popularity in various fields as a versatile fabrication tool to create complicated shapes. There have been a number of studies to design and 3D-print shock-absorbing structures. \cite{Tsouknidas2016} have evaluated impact absorption of 3D-printed porous polylactic acid (PLA) structures. \cite{Bates2016} have investigated energy absorption of 3D-printed honeycomb structures made of thermoplastic polyurethane (TPU). Recently, \cite{Chen2018} have analyzed crushing behavior of graded lattice cylinders, 3D-printed using Acrylonitrile Butadiene Styrene (ABS) plastic, subject to axial impact. 

In this study, we evaluate nonlinear wave dynamics in 3D-printed, graded lattices composed of hollow elliptical cylinders (HECs). The gradient in the geometry of the HECs along the chain results in the asymmetric dynamics of stress wave propagation. That is, the impact mitigating behavior in the positive gradient chain and that in the negative gradient chain are highly distinctive. The nonreciprocal wave propagation in nonlinear systems has been studied extensively, mainly relying on the multi-stability of their unit cells. 
For example,~\cite{Nadkarni2016} have found a mechanical diode using bistable lattices. \cite{Fang2017} have used multi-stable stacked-origami to realize a static mechanical diode effect. \cite{Wu2018,Wu2018arXiv} have recently reported unidirectional transmission of energy in metastable structures. 
However, the asymmetric dynamics in a nonlinear mechanical system without relying on multi-stability has been relatively unexplored.

Here, we numerically and experimentally verify the impact absorbing mechanism of graded HEC structures in 1D as well as 2D architectures. In particular, we leverage the strain-softening nature of the HECs to demonstrate the asymmetric wave dynamics. The stiffness and the mass gradient combined with the strain-softening nonlinearity create the unique dynamics.
These HEC lattices can be easily fabricated using 3D-printing and assembled by simple bonding, in sharp contrast to granular crystals which need delicate mechanical contact. This provides us the following advantages: First, we can avoid the issues regarding the energy dissipation due to the rotational and shear friction of granular beads in experiment. Second, we have the flexibility to design and the convenience to fabricate the unit cells. These advantages enable us to customize the unit cells to show various behaviors and investigate their effect on wave dynamics. Finally, we can achieve a low weight-to-volume ratio structure by using generic polymer materials as well as by optimally designing the structures. 

The rest of the manuscript is arranged in the following manner. In Section~\ref{Sec:method}, we provide details on the technical approach. Section~\ref{Result:asymm} describes the asymmetric nonlinear dynamics in the HEC chain with a positive and a negative gradient direction. We experimentally demonstrate the simulated dynamics in the graded HEC chains. In Section~\ref{Result:Impact}, we conduct rigorous comparison of the impact mitigation performance between the two opposite gradient chains. Section~\ref{Result:2D} suggests a potential extension of the 1D graded HEC chains to the 2D graded HEC array. We finish our manuscript with conclusion in Section~\ref{Conclusion}.

\section{Methods}\label{Sec:method}

\begin{figure}[th]
 \begin{center}
 \includegraphics[width=1\linewidth]{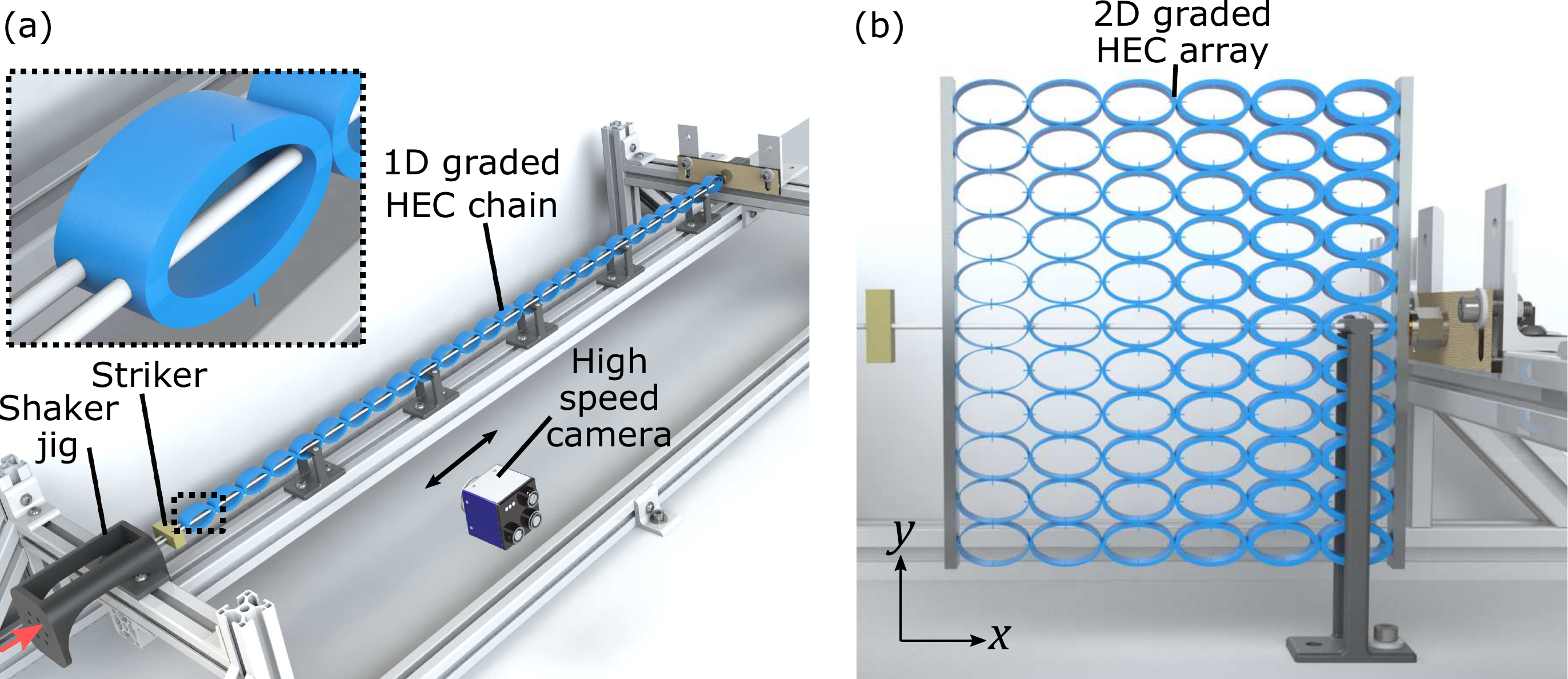}
 \end{center}
 \caption{(a) A schematic diagram of the 1D HEC setup for dynamic testing. Shaker jig is attached to the shaker head (not shown but indicated by the red arrow) to slide the striker. The graded HEC chain is impacted by the striker on the left end at the input velocity, $v_s=2.95 \pm 0.09$ m/s. The right end is in contact with a piezoelectric sensor. The inset is a close-up view of the single HEC inside the dotted box. (b) A schematic diagram of experiment setup for the 2D graded HEC array.}
 \label{fig:exp_setup}
\end{figure}

We first perform 1D experiment as follows: we 3D-print 26 HECs using PLA material with their wall thicknesses linearly varying from 0.4 mm to 3 mm. 
See \ref{ApdxB} for the detailed geometry and force-displacement relationship of the HECs under the variation of their wall thicknesses. 
We align these HECs into two different chain configurations: (i) a chain with the increasing HEC thicknesses (called a thickening chain for the rest of this article) and (ii) a chain with the decreasing HEC thicknesses (i.e., thinning chain). Figure~\ref{fig:exp_setup}(a) shows an experimental setup of the thinning chain. Here, the HECs are printed with two small holes (diameter of 2.50 mm) along the major axis, such that they are allowed to slide along the horizontal direction by using a pair of stainless steel rods (diameter of 2.38 mm) that penetrate through all HECs (inset of Fig.~\ref{fig:exp_setup}(a)). These rods are held by 3D-printed supports to prevent them from bending. To ensure their mechanical contact, the HECs are bonded together using glue (Loctite 431). 

For the application of mechanical impact, we send an impulse signal to a shaker (LDS V406, B\&K) such that it launches a rectangular striker mass towards the one end of the HEC chain at a desired velocity. The other end of the chain is bounded by a piezoelectric sensor (208C02, PCB Piezotronics), which is mounted on the rigid chassis and is used to measure the transmitted forces through the HEC chain. The striker mass ($m_s=4.3 $ g) is 10 times larger than the mass ($m$) of the thinnest cylinder. When the striker hits the chain, it triggers the high speed camera (Phantom v1211) to start recording. The high speed camera moves along the linear stage, capturing three cylinders at a time at 40000 fps. The measurements are made at every other cylinder location, and we iterate the experiment five times for statistical treatments. To facilitate the tracking of the HECs' translational motions, we print bosses on the top and bottom of each HEC (see the inset of Fig.~\ref{fig:exp_setup}(a) or Fig. A.1(a) in \ref{ApdxB}). 
We use the digital image correlation technique to measure the displacements of HECs~\citep{HKim2018}. 
The data collected are stitched together, and the aligned data after reconstruction in the space domain are presented in the results hereafter.

To compare with the experimental results, we perform numerical analysis using commercial finite element analysis (FEA) software (ABAQUS). We apply the Timoshenko beam model (B22 element) to capture the behavior of the HECs. 
To take into account the hyperelastic behavior of PLA, we implement the Neo-Hookean model. We find its shear modulus and material damping factor by fitting the experimental data based on uniaxial compression tests. 

We also compose a 2D graded HEC array as shown in Fig.~\ref{fig:exp_setup}(b). We print 11 rows of HECs whose thicknesses vary linearly from 0.4 mm to 3 mm along the horizontal direction. Then we carefully align them and glue between the rows into a $6\times 11$ array. Similar to the 1D setup, the HECs in the middle row are guided by a pair of stainless steel shafts. We cut two 5-mm-thick aluminum plates to a size of 200 mm by 12 mm with two small holes (diameter of 2.5 mm) and glue them to the left and the right end of the graded HEC array. These plates and HEC arrays are equivalent to the facesheet and core of sandwich structures, respectively, in 2D representation. The right plate is in contact with the piezoelectric sensor to record the transmitted forces through this 2D HEC array. To prevent any rotational dynamics of the array, we firmly hold the horizontal guiding rods by using a 3D-printed support. The way we apply striker input is the same as that in the 1D chain experiment.

\section{Results and Discussion}\label{Sec:Result}
\subsection{Asymmetric impact response}\label{Result:asymm}

\begin{figure}
 \begin{center}
 \includegraphics[width=1\linewidth]{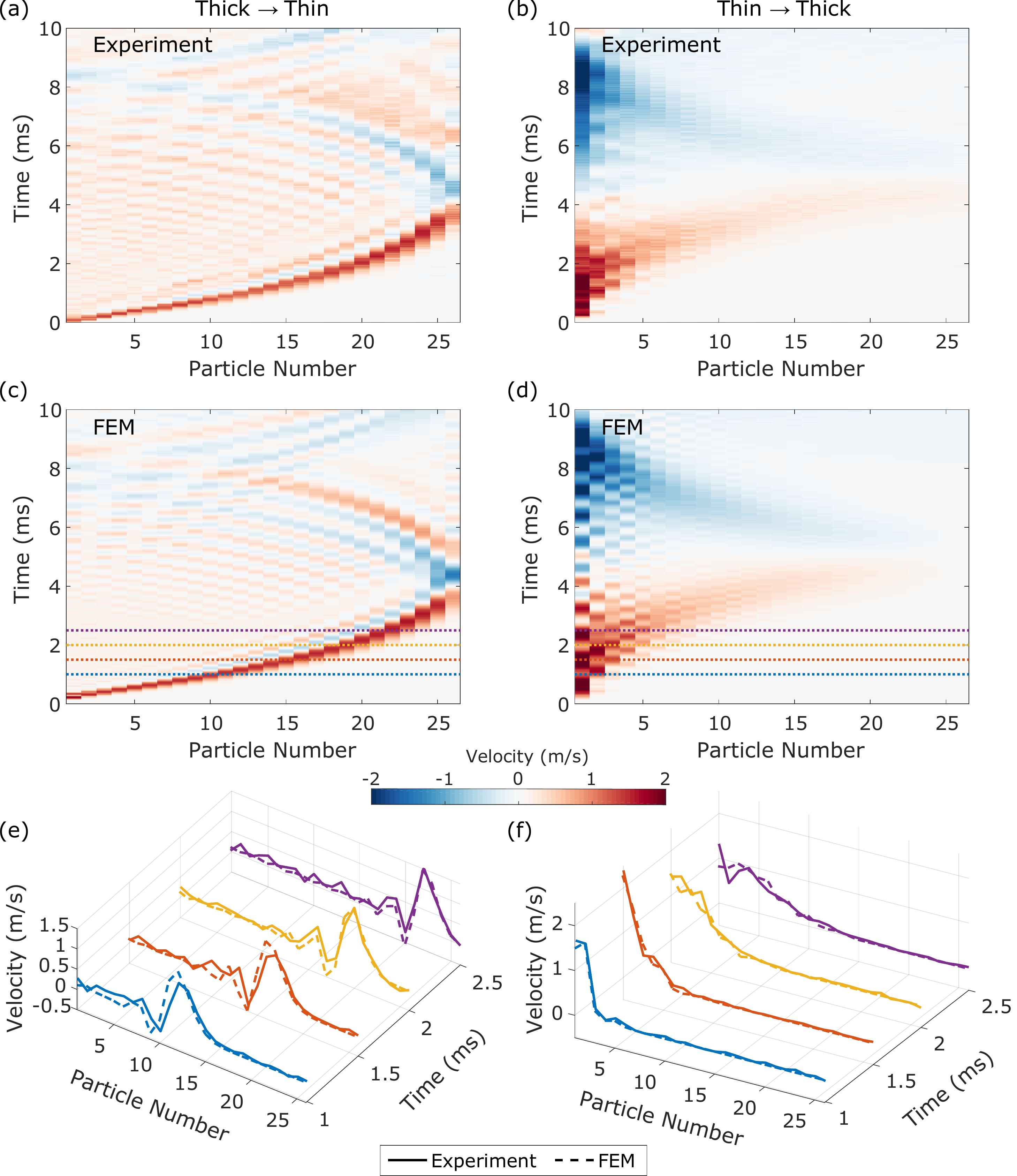}
 \end{center}
 \caption{Surface map of HECs' velocity profiles in space and time domain for (a) the thinning and  (b) the thickening chain, based on the experimental data. The FEA results for the corresponding configurations are shown in (c) and (d). The cross sections along the dotted lines in (c) and (d) are plotted in (e) and (f) in dashed curves, respectively. The solid curves represent the corresponding velocity profiles from the experimental data.
}
 \label{fig:vel_surf}
\end{figure}

Figure~\ref{fig:vel_surf} shows experimental (top row) and numerical (middle) results of HECs' velocity profiles for thinning (left) and thickening (right) chains.  
It is notable that the thinning and the thickening chain show completely different dynamic behaviors. The first feature to notice is the acceleration/deceleration of the propagating waves depending on the gradient direction. This phenomenon is closely related to the mass and stiffness of the unit cells composing the chain. 
In case of the thinning chain, the contact stiffness decreases towards the end of the chain. Lower stiffness results in the decrease in wave speed (Fig. \ref{fig:vel_surf}(a) and (c)). Furthermore, the negative mass gradient will amplify the propagating wave amplitude, slowing down the wave due to the effective strain-softening nature of the HECs. 
In the same manner, acceleration occurs in the opposite direction (Fig. \ref{fig:vel_surf}(b) and (d)). The experimental data (Fig.~\ref{fig:vel_surf}(a) and (b)) corroborate the FEA results (Fig.~\ref{fig:vel_surf}(c) and (d)) for both thinning and thickening chains.

The accelerating and decelerating waves in graded chains have been also explored in previous studies \citep{Sen2001,Melo2006,Doney2009,Harbola2009Tap,Machado2013,Chaunsali2018}. However, our system is different from those in previous studies in that (1) the unit cells are bonded together supporting both compressive and tensile motions, and (2) they are deformed modestly in a weakly nonlinear regime. As a result, we observe unique wave dynamics in the HEC system, such as the formation of weakly nonlinear pulses followed by oscillatory tails due to tensile motions (see the ripples in Fig.~\ref{fig:vel_surf}).

Another feature to notice is the amplitude change. The amplitude significantly decreases as the wave approaches the end of the thickening chain (Fig. \ref{fig:vel_surf}(b) and (d)). However, in the thinning case, the amplitude increases minutely as the wave propagates along the chain (Fig. \ref{fig:vel_surf}(a) and (c)). This amplitude change is obvious if we look at the individual wave profiles in Fig. \ref{fig:vel_surf}(e) and (f). This attenuation/amplification of the propagating wave can be also explained by the change in mass and stiffness within the graded chain. 
In the thickening chain, the mass of the HEC increases, thereby decreasing the particle velocity to satisfy the conservation of momentum principle. Moreover, the increasing stiffness results in a faster group velocity, which extends the width of the leading wave and contributes to the attenuation of the wave amplitude. On the contrary, in the thinning chain, the HEC masses become lighter towards the end, resulting in higher particle velocities. Contact stiffness also decreases in this case, which contributes to narrowing width and thus amplifying the wave. More details on the effect of the contact stiffness and the mass are described in \ref{ApdxE}. The amplitude change also is affected by the energy transfer from kinetic energy to strain energy. See \ref{ApdxD} for the energy analysis.

The asymmetric wave propagation in opposite directions is in reminiscence of the study by \cite{Wu2018,Wu2018arXiv}. They take advantage of the asymmetry of their system to realize non-reciprocal wave propagation by selectively triggering nonlinear instability in a single direction. The transmission induced by overcoming the threshold with high amplitude input is called supra-transmission. Similarly, for the thinning chain in our system, the striker impact contains a wide range of frequencies (Fig.~\ref{fig:FFT}(a) in \ref{ApdxF}) and the output signal contains most of the input frequency components, despite the significant transfer to the low frequency modes. The thickening chain does not transfer much energy to the end of the chain, as seen in Fig.~\ref{fig:FFT}(b) in \ref{ApdxF}. As a result, we confirm a similar trend of asymmetric supra-transmission in our graded HEC chain, by leveraging not the bistability of unit cells, but their arrangements in gradient. More details on the spectral analysis can be found in~\ref{ApdxF}.

\subsection{Impact mitigation}\label{Result:Impact}
%

\begin{figure}
\centering
 \includegraphics[width=1\linewidth]{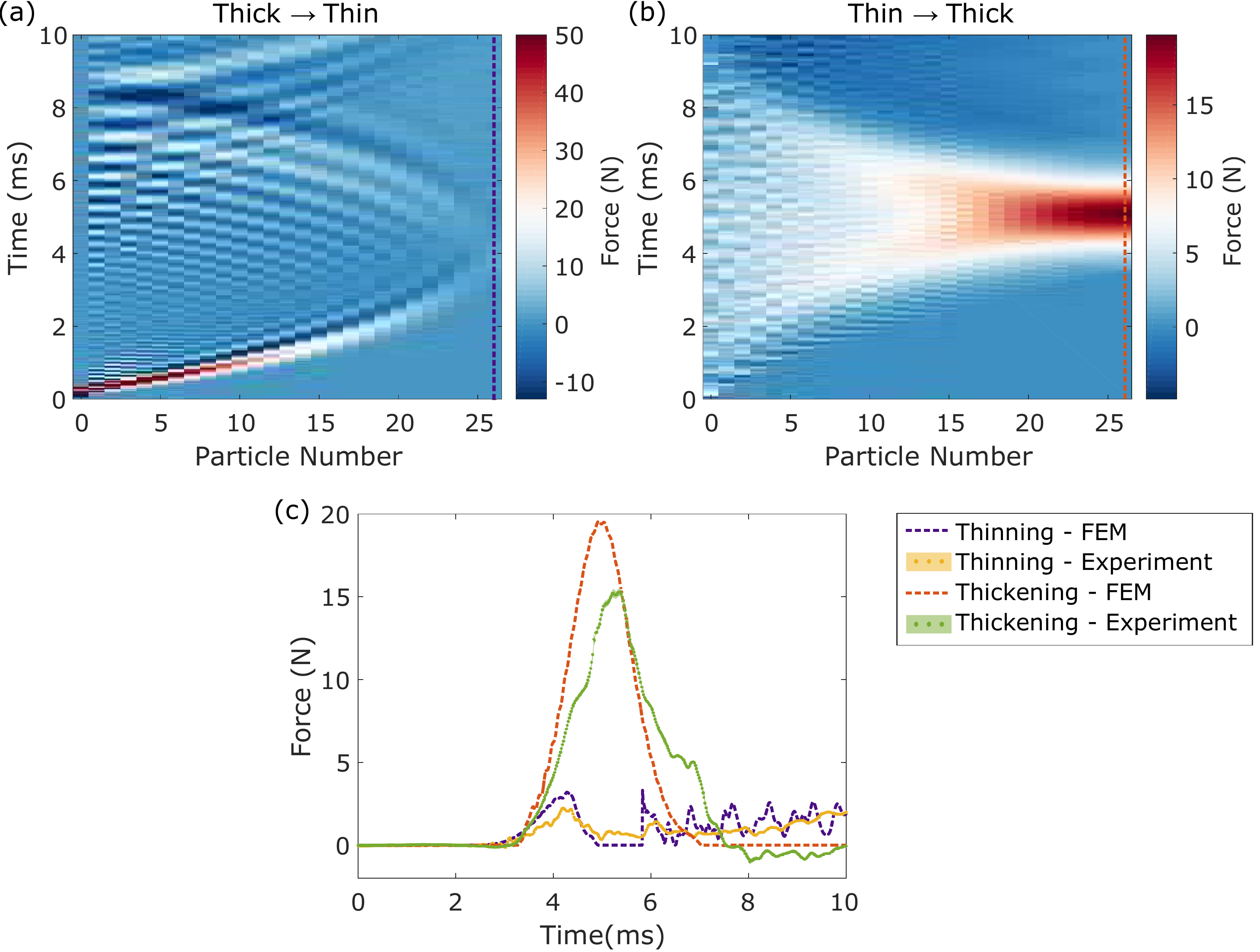}
 \caption{Contact force between HECs in (a) the thinning chain and (b) the thickening chain. The contact force is calculated in ABAQUS simulations. 0 index means the striker. Negative force means tensile direction. The dashed lines at $n=26$ indicate the output force profiles which are plotted in (c). (c) Contact force output at the right boundary. The purple and red dashed curves show the output force in the thinning and the thickening HEC chain from FEA results, respectively. We run 10 experiments and plot the average values in circular markers and the standard deviations in shaded areas (yellow for the thinning and green for the thickening HEC chain). The striker velocity $v_s=3.00\pm0.07$ m/s).
}
 \label{fig:CF}
\end{figure}

We evaluate the shock absorption performance of the graded HEC chains by investigating their transmitted force profiles (Fig.~\ref{fig:CF}). In the thinning chain, the initially localized force peak disperses slowly as the wave propagates along the chain (Fig.~\ref{fig:CF}(a)). In the thickening chain, the initially dispersed force profile tends to become highly localized towards the end of the chain (Fig.~\ref{fig:CF}(b)). 
This difference is evident if we compare the force profiles measured at the end of the chain by using a force sensor (Fig.~\ref{fig:CF}(c)). Here, we also include experimental results for comparison (circular markers). We observe a solitary wave-like localized pulse in the thickening chain (see the dashed red curve based on the FEM result, which show no oscillatory tails). On the other hand, the thinning chain exhibits a small-amplitude leading pulse followed by oscillations. While there exist noticeable discrepancies between the experimental and computational results possibly due to friction and material damping, we still witness the qualitative difference in the empirical force packets between the thinning and thickening chains.

\begin{figure}[t]
\centering
 \includegraphics[width=1\linewidth]{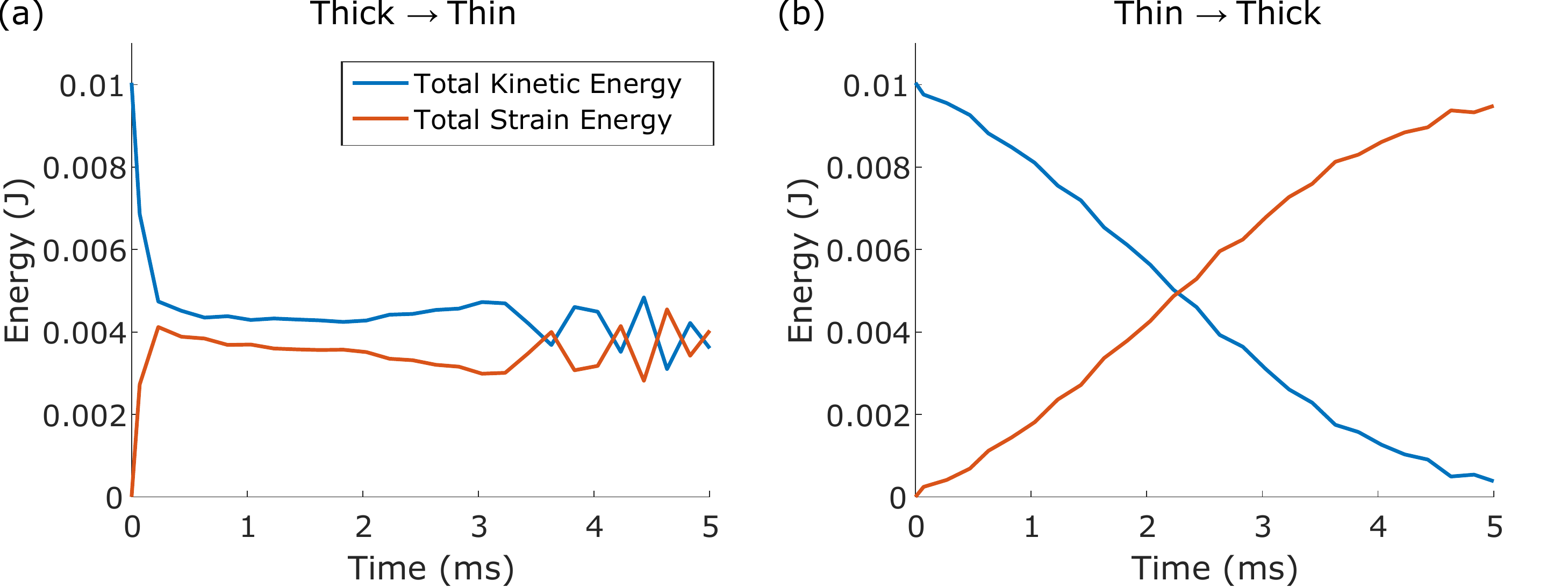}
 \caption{Kinetic energy (blue curve) and strain energy (orange curve) of the entire model in (a) the thinning chain and (b) the thickening chain based on numerical simulations without damping. 
}
 \label{fig:Eall}
\end{figure}

These force profiles may look contradictory to the velocity profiles presented in Section~\ref{Result:asymm}. In the velocity profiles, the thinning chain showed the focusing behavior of wave packets (left column in Fig. \ref{fig:vel_surf}), while the thickening chain exhibited the attenuation of those packets (right column in Fig. \ref{fig:vel_surf}). This seemingly contradictory behavior between velocity and force profiles can be explained by the energy analysis. We calculate the evolution of kinetic and strain energy stored in the HEC system over the span of 5 ms, which covers a travel of the initial impact to the end of the chain (Fig.~\ref{fig:Eall}). In the thinning chain, we observe the kinetic energy and the strain energy reach an equilibrium point where their levels are comparable (Fig.~\ref{fig:Eall}(a)). In the thickening chain, however, most kinetic energy is converted into the strain energy. In other words, in the thinning chain, the impulse facilitates the rattling of the HECs after it passes by the chain (thus causing the oscillatory tails as witnessed in Fig.~\ref{fig:CF}(c)), while the thickening chain rather piles up its energy into the strain energy instead of shedding it into dynamic energy. This is the key difference in energy transfer between the thinning and thickening chains. Since the velocity profiles are directly related to the kinetic energy, the trend looked contradictory to the force profiles that are associated with the strain energy. Further details on energy analysis can be found in~\ref{ApdxD}.

\subsection{Two-dimensional expansion}\label{Result:2D}
\begin{figure}[t]
 \begin{center}
 \includegraphics[width=1\linewidth]{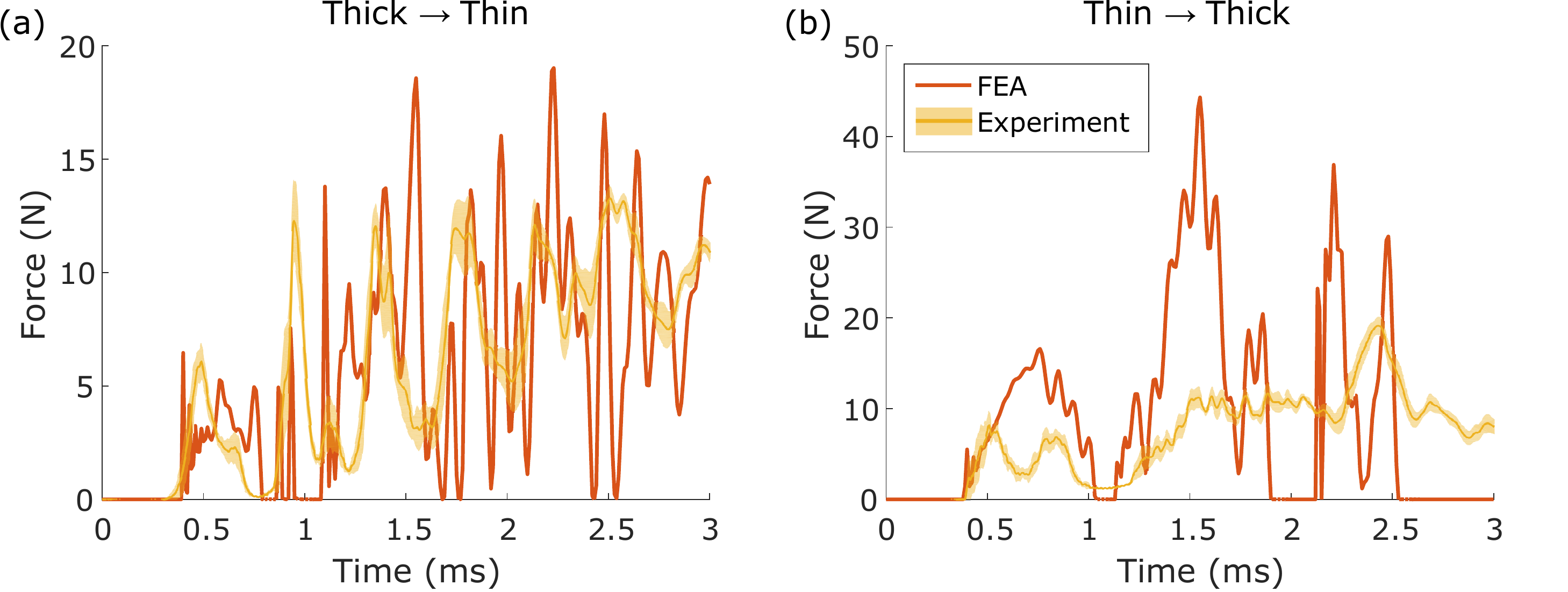}
 \end{center}
 \caption{Contact forces at the output boundary are plotted for (a) the 2D thinning array  and (b) the 2D thickening array. The array size is $6\times 11$. We run 10 experiments and plot the average in the yellow solid line and the standard deviations in the yellow shaded areas ($v_s=2.93\pm0.05$ m/s).
}
 \label{fig:CF_2D}
\end{figure}

Now we assess the feasibility of extending the 1D HEC system into a 2D one, such that it can be potentially used as a core material for impact mitigating structures. To this end, we build a sandwich structure as shown in Fig. \ref{fig:exp_setup}(b). Here, the arrangement is in a pseudo-2D manner, graded in the $x$ direction and homogeneous in the $y$ direction. Given the shallow thickness of the layer (12 mm), we impose the plane stress condition in the out-of-plane direction. We measure the transmitted force profiles at the right end of the prototype and plot them in yellow curves for the thinning (Fig. \ref{fig:CF_2D}(a)) and the thickening (Fig. \ref{fig:CF_2D}(b)) array. The corresponding numerical data are also plotted in red curves in Fig. \ref{fig:CF_2D}. Consistent with the 1D chain result, the maximum force at the wall in the thinning array is lower than that of the thickening chain. However, the performance of the thinning chain is better than the thickening chain only by twice, which is smaller than the 7-fold improvement observed in the 1D system. This is likely due to the transverse dispersion of energy, as well as the short chain length in 2D. 

\begin{figure}[t]
\centering
 \includegraphics[width=1\linewidth]{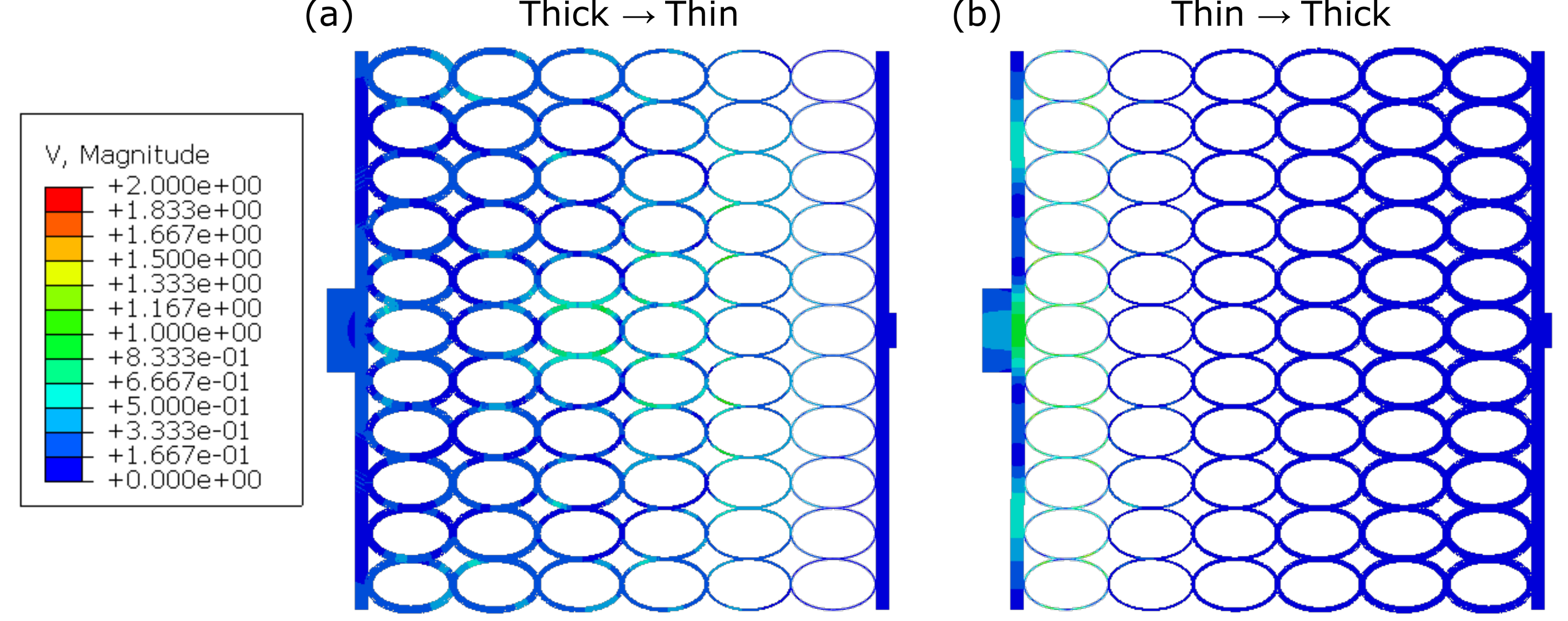}
 \caption{Velocity contour of the 2D model from ABAQUS for (a) thinning HEC and (b) thickening HEC array at $t=0.4$ ms.
}
 \label{fig:2Dvel}
\end{figure}

The dispersive effect can be clearly seen in numerical simulations (Fig.~\ref{fig:2Dvel}), where the thickening lattice immediately radiates energy in the transverse direction (\textit{y}-direction) as the striker hits the plate, while the thinning lattice does not. Despite this transverse dispersion effect, we still verify the asymmetric wave propagation in the 2D graded HEC array. Particularly, the amplifying velocity towards the end of the thinning array is consistent with the dynamics we observe in the 1D thinning chain (Fig.~\ref{fig:vel_surf}(a)). Likewise, the localization of the velocity at the input boundary of the thickening array is in agreement with the 1D thickening chain (Fig.~\ref{fig:vel_surf}(b)). 

\begin{figure}[t]
 \begin{center}
 \includegraphics[width=1\linewidth]{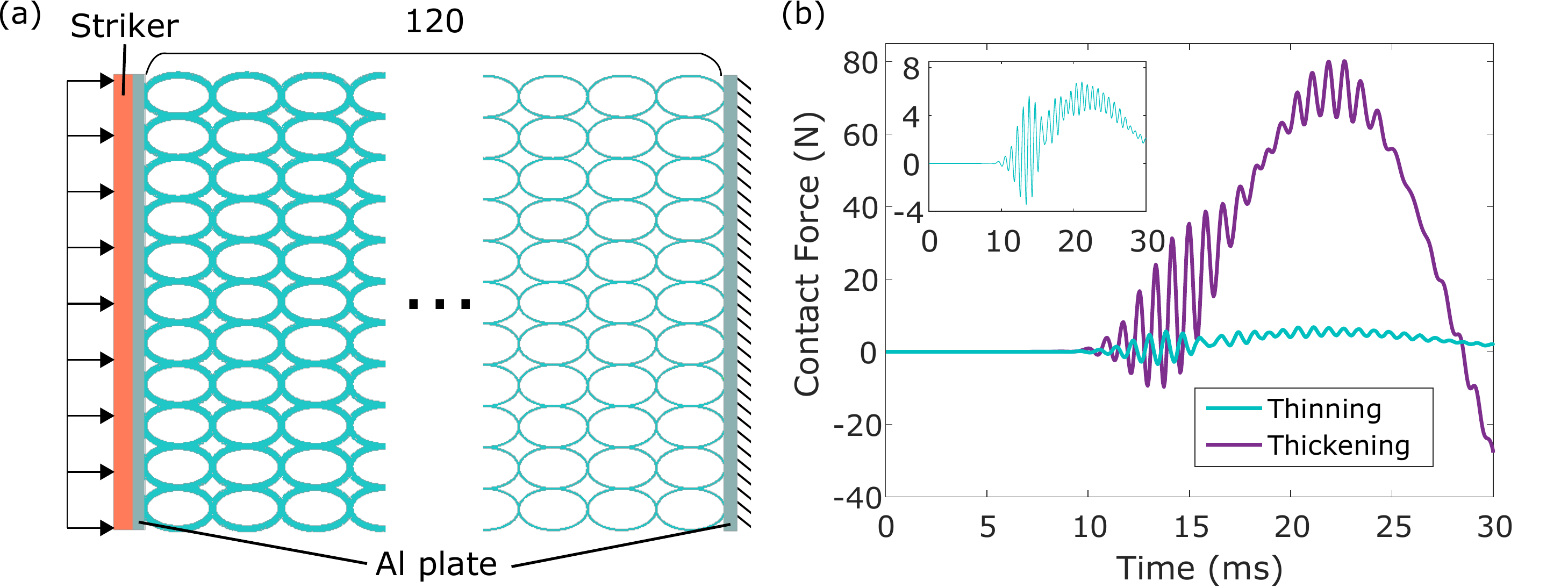}
 \end{center}
 \caption{(a) The schematic of the $120\times 11$ HEC model. The thickness gradient is assigned such that it varies from 3 mm to 0.4 mm along the impact direction. The 2D HEC core is bonded and sandwiched between aluminum plates. The plate thickness is determined such that its mass is the same as the thickest HEC column's mass. (b) FEA analysis on contact force at the right end for 2D thinning array (turquoise line) and 2D thickening array (purple line). The inset shows the thinning chain response in a magnified view. Damping is not considered in these simulations.
}
 \label{fig:CF_2D_L}
\end{figure}

To investigate the chain size effect, we model a larger array, $120\times 11$, as shown in Fig.~\ref{fig:CF_2D_L}(a) and examine how the impact absorbing level changes. A striker with the mass of 30 g is applied at an initial velocity of 3 m/s. We observe an order-of-magnitude difference in the maximum output force between the thinning and the thickening arrays (Fig.~\ref{fig:CF_2D_L}(b)). The oscillating envelope of the force profile is due to the vibrations of the HECs. In fact, the dominant frequency component of the wave profile is 1.2 kHz, which is around the $1^{st}$ vibration mode of the thickest HECs and the $2^{nd}$ vibration mode of the thinnest HECs. This numerical result hints that the proposed 2D HEC lattices can be potentially used as a core material in sandwich structures for incurring asymmetric wave propagation and thus efficient impact mitigation.

\section{Conclusion}\label{Conclusion}
We investigated nonlinear dynamic behavior of the graded lattice made of hollow elliptical cylinders (HECs). We experimentally and numerically demonstrated the asymmetric wave dynamics under the condition of external impact applied in opposite directions. This includes acceleration/deceleration and attenuation/amplification of the mechanical wave in terms of velocities, depending on whether the system has a positive or negative gradient of mass and stiffness imposed on the HEC chain. We also investigated the force transmission through the one-dimensional chain, whose trend is seemingly flipped with respect to the velocity profiles. We explained this contradictory behavior by energy analysis. We extended our findings from the 1D system to the 2D array, and demonstrated the efficacy of this asymmetric wave dynamics for impact mitigation purposes. This expansion in 2D space sheds light on a new way to design sandwich core structures for controllable stress wave management. Assigning stiffness gradients in both directions of the 2D array is an on-going study and will potentially lead to the manipulation of wave directionality. 

\section*{Acknowledgments}  
J.Y. and H.K. thank the support of the National Science Foundation under Grant No. CAREER-1553202. E.K acknowledges the support from the National Research Foundation of Korea (NRF) grant funded by the Korea government (MSIP) under Grant No.2017R1C1B5018136

\clearpage
\appendix
\raggedbottom\sloppy
\counterwithin{figure}{section}
\renewcommand*\thefigure{\Alph{section}.\arabic{figure}}

\section{Force-displacement relationship of HECs}\label{ApdxB}
\begin{figure}[ht]
\centering
 \includegraphics[width=0.5\linewidth]{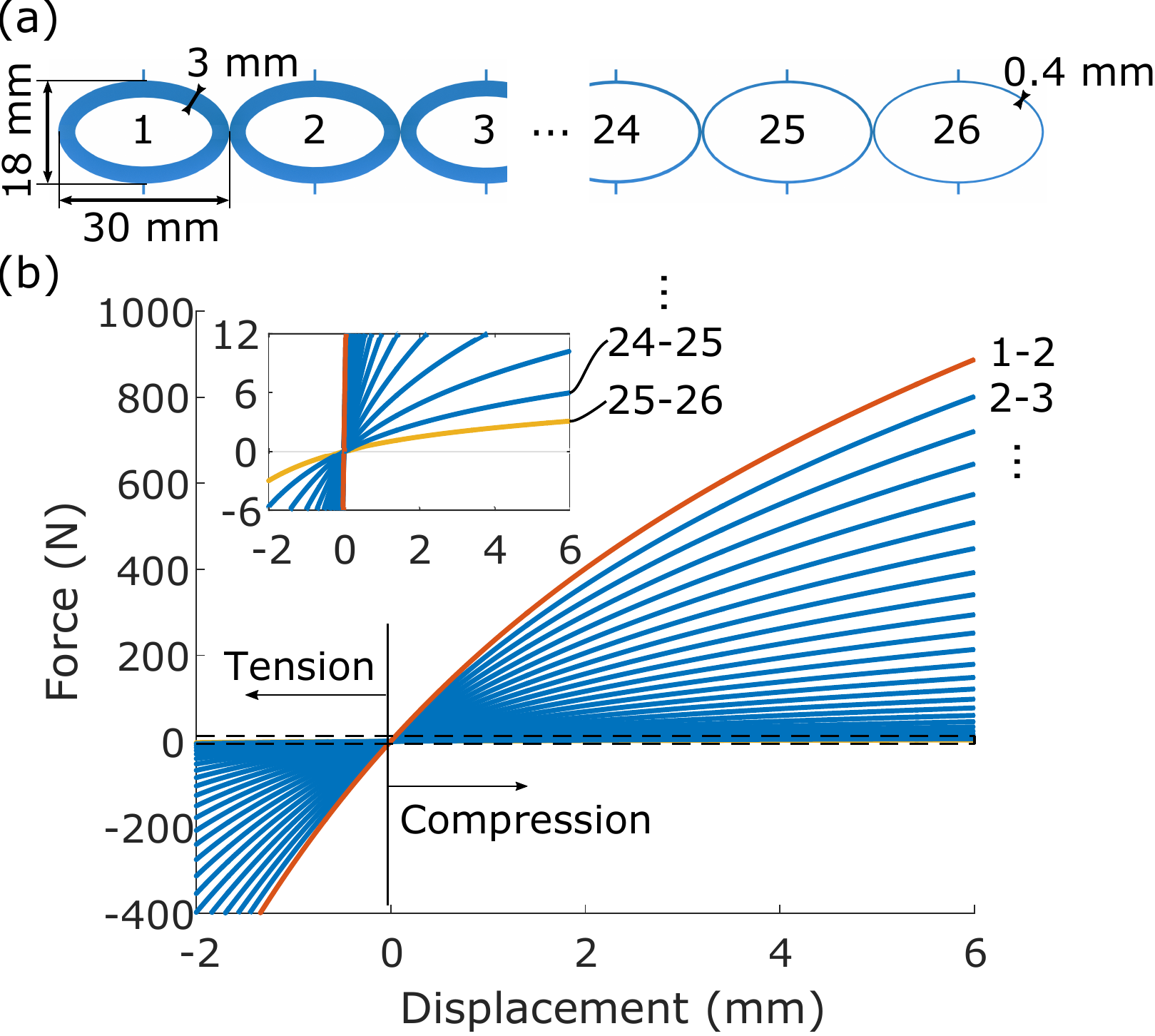}
 \caption{(a) An illustration of the graded HEC chain and its dimensions. The chain is composed of $N=26$ HECs with linearly graded thickness from 3 mm to 0.4 mm. The out-of-plane width is 12 mm. (b) The force-displacement curve between adjacent HECs in the graded chain. The compressive deformation is plotted positive. The force-displacement relationship of cylinder 1 and 2 is plotted in the orange curve. Likewise, the force-displacement relationship between cylinder 2 and 3, 3 and 4, and so on (refer to (a) for cylinder numbers) are plotted respectively from the top to the bottom. The inset shows an enlarged view of the force-displacement curves for the last few cylinder pairs.
} \label{fig:F-d}
\end{figure}

In this study, we achieve the variations of the HECs' stiffness by modifying their wall thicknesses. To assess this thickness effect, we numerically calculate the force-displacement relationship between adjacent cylinders in the graded HEC chain using FEA (Fig.~\ref{fig:F-d}(a) for the geometrical configuration of the gradient chain composed of HECs). We see from Fig.~\ref{fig:F-d}(b) that the HECs show effective strain-softening behavior, i.e., decreasing stiffness (the slope of the force-displacement curve) in compression. This holds true regardless of the thickness of the HECs, as seen from the force-displacement curve of thickest pair (the orange curve in Fig.~\ref{fig:F-d}(b)) to the thinnest pair of HECs (the yellow curve in the inset of Fig.~\ref{fig:F-d}(b)). In fact, the force-displacement curve of the HEC nearly scales up as its thickness increases. As a result, the graded HEC chain exhibits a gradual change of force-displacement relationship, leading to a universal power-law relationship~\citep{HKim2018}. 

\section{Amplitude-dependent wave speed}\label{ApdxE}
\begin{figure}[th]
 \begin{center}
 \includegraphics[width=1\linewidth]{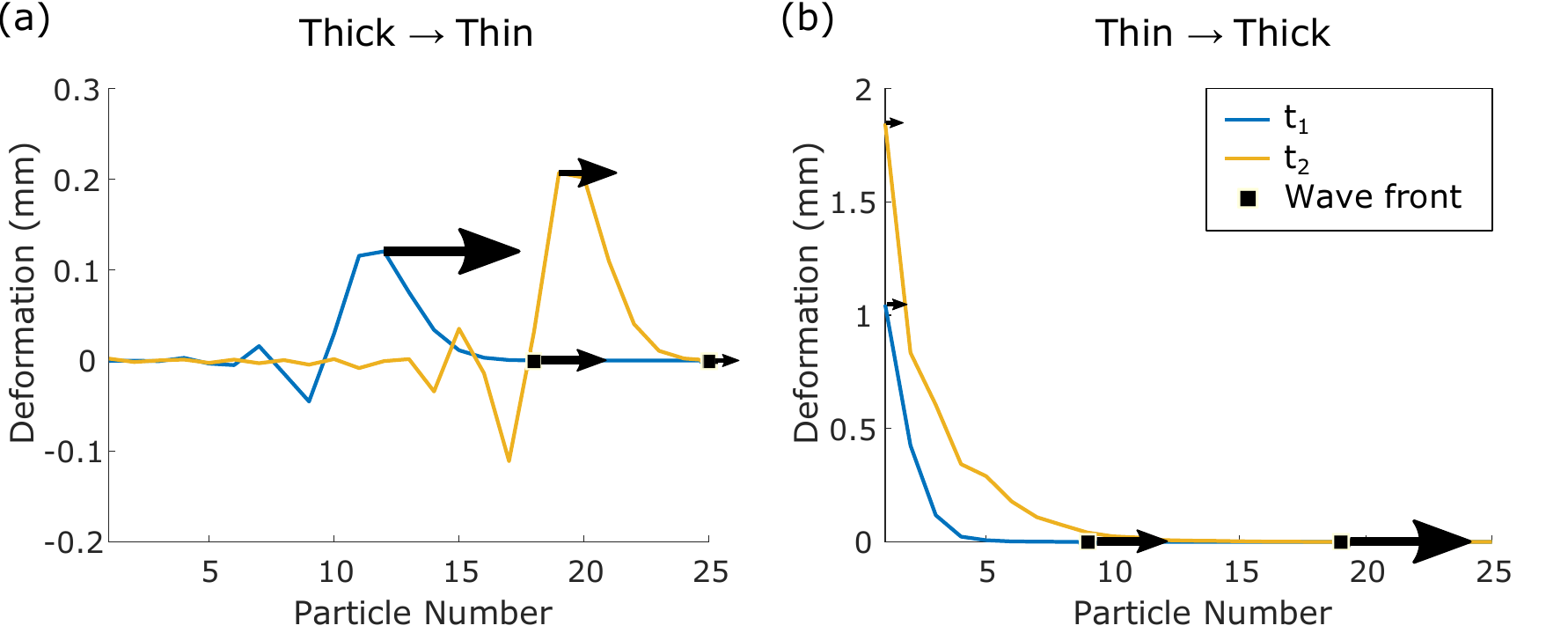}
 \end{center}
 \caption{Snapshot of deformation profiles at two time points in (a) the thinning and (b) the thickening HEC chain. The blue curve is at time 1 ($t_1$) and the yellow curve is at time 2 ($t_2$) where $t_1<t_2$. The arrows indicate wave speed of the corresponding points. The length of the arrow is relative scale of the wave speed.}
 \label{fig:1D_cg}
\end{figure}

In this section, we qualitatively explain why the amplitude-dependent wave speed deforms the shape of the leading wave. We assume the wave speed is proportional to the stiffness and inversely-proportional to the mass, i.e., $c_g\propto\sqrt{k/m}$ where $c_g$ is the wave speed, $k$ is the stiffness, and $m$ is the mass. Since both the stiffness and the mass vary together in our graded HEC chain, we need to consider both to estimate the wave speed. 
Figure~\ref{fig:1D_cg} shows how deformation wave profile changes as the wave propagates through the chain. For the thinning chain, as shown in Fig.~\ref{fig:1D_cg}(a), the leading wave stiffens. Although the later part of the chain has smaller masses, the stiffness drops much more drastically. As a result, the thinning chain has positive relation between the amplitude and the wave speed (wave peak travels faster than the wave front). On the contrary, we observe expansion of the leading wave in the thickening chain (Fig.~\ref{fig:1D_cg}(b)). The increase in stiffness overshadows the increase in mass towards the chain end, resulting in negative relationship between the amplitude and the wave speed. 

\section{Energy analysis}\label{ApdxD}

\begin{figure}[th]
\centering
 \includegraphics[width=1\linewidth]{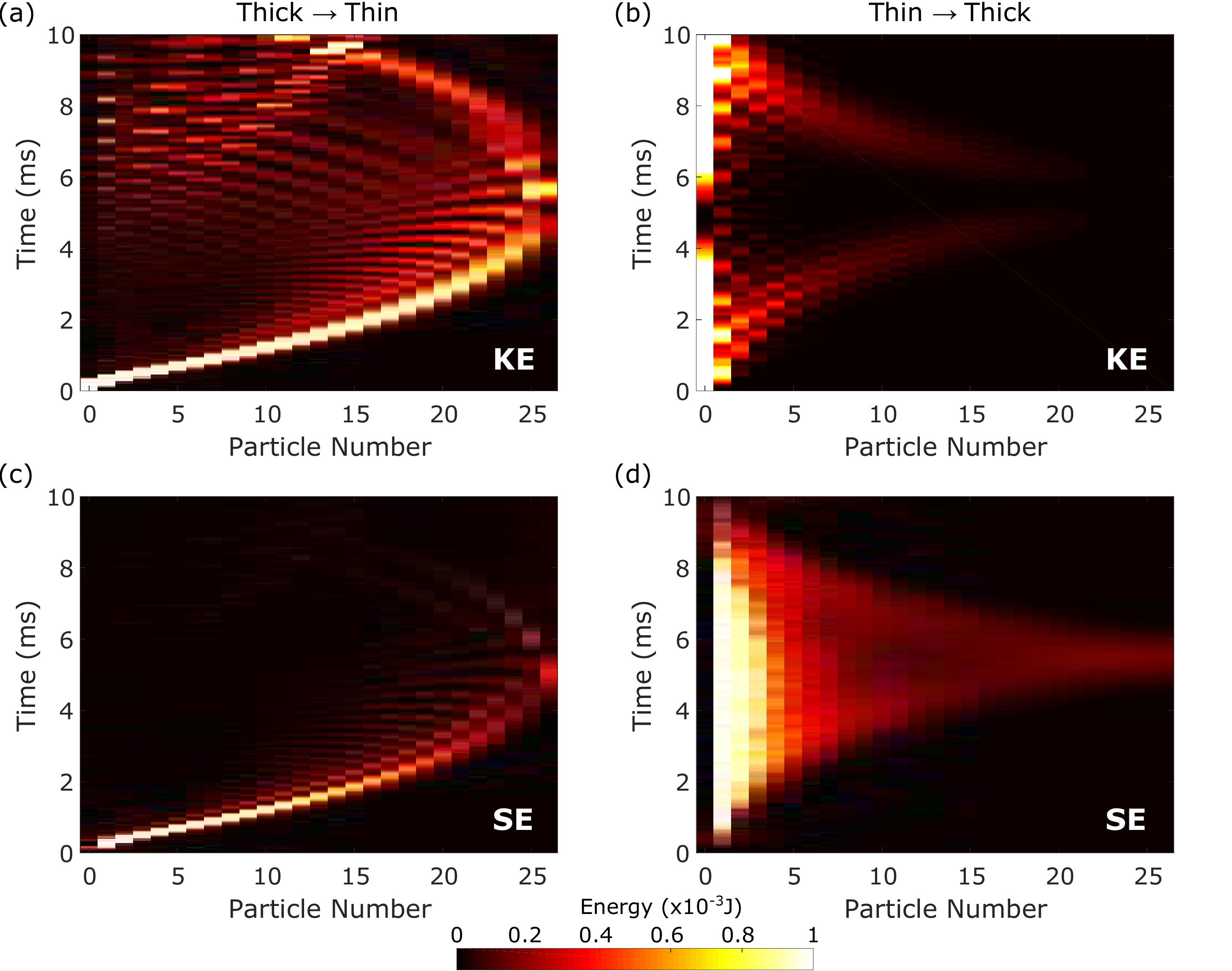}
 \caption{The evolution of kinetic energy (KE) and strain energy (SE) is plotted for thinning chain in (a) and (c) and for thickening chain in (b) and (d), respectively. The energy is calculated in ABAQUS simulations without damping. 
 0 index means the striker.
}
 \label{fig:Energy}
\end{figure}

In this section, we investigate the evolution of kinetic and strain energy during the propagation of waves in thinning and thickening HEC chains. We export kinetic and strain energy data from ABAQUS simulation results and plot the contour map in both time and space domain in Fig.~\ref{fig:Energy}.
For the thinning chain, we see a large portion of the kinetic energy (Fig.~\ref{fig:Energy}(a)) is transmitted towards the end of the chain, even showing clear reflection from the boundary. An equivalent amount of the strain energy (Fig.~\ref{fig:Energy}(c)) is also observed, but it diminishes noticeably as the wave propagates through the chain. 
The thickening chain shows a different trend of kinetic and strain energy transmission. We see a drastic reduction of the kinetic energy in the first few HECs in the chain (Fig.~\ref{fig:Energy}(b)). This kinetic energy lost is transferred to the strain energy as shown by the highlighted area in Fig.~\ref{fig:Energy}(d). This energy transfer phenomenon is again due to the large deformation of the thin HECs in the early part of the chain, caused by the striker impact. This corroborates the results reported in the velocity and force profiles as shown in Fig.~\ref{fig:vel_surf} and \ref{fig:CF}.

\section{Spectrum interpretation}\label{ApdxF}
\begin{figure}[t]
\centering
 \includegraphics[width=1\linewidth]{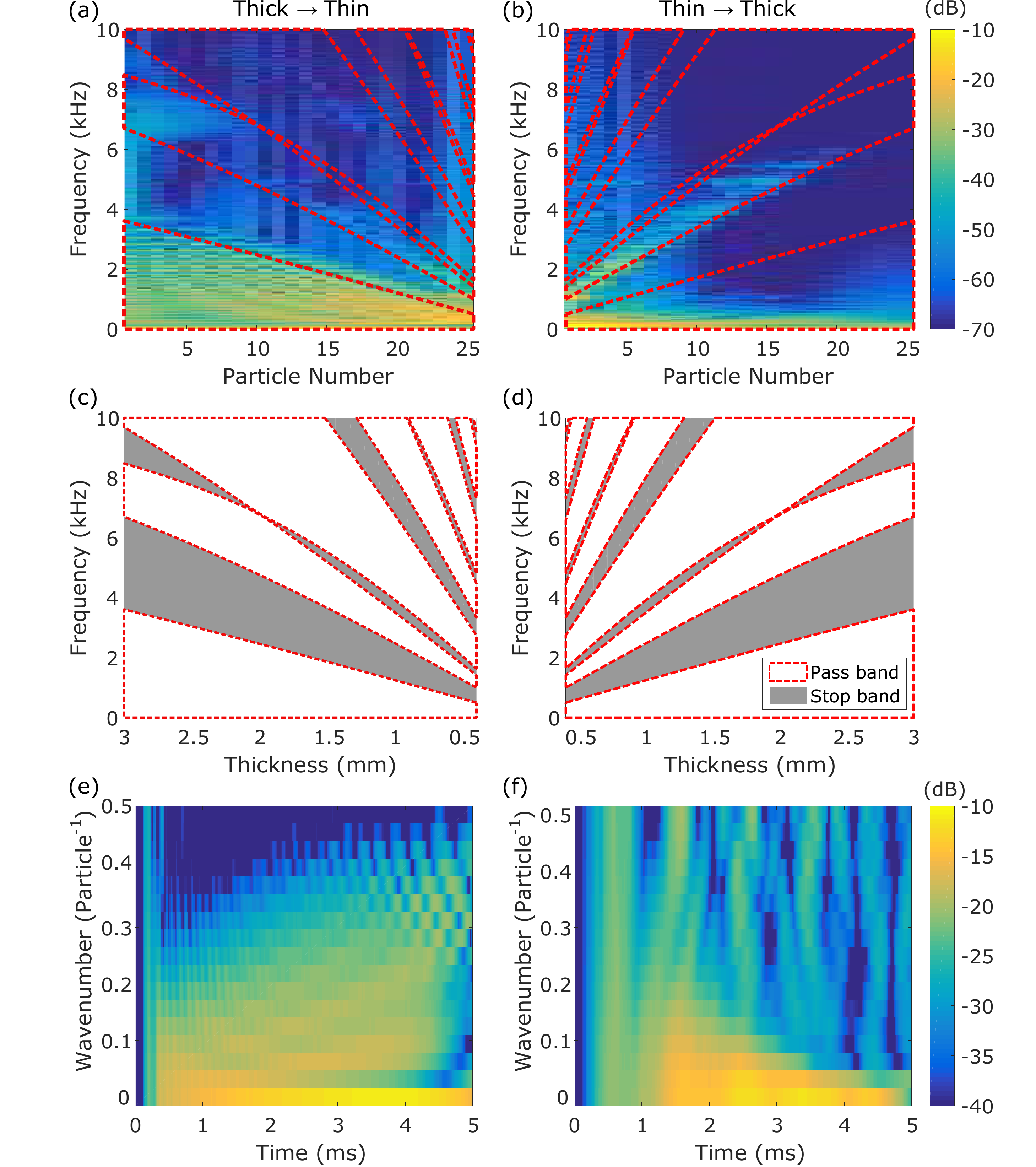}
 \caption{Spectral contour of the velocity data from FEA without damping is calculated using Fast Fourier Transform (FFT) in time domain for (a) the thinning and (b) the thickening chain. The red dashed boxes represent the pass bands obtained from the band structures in (c) and (d) for an infinite thinning and thickening chain, respectively. Spatial spectrum of the velocity data is also calculated using FFT in space domain for (e) the thinning and (f) the thickening chain up to 5 ms around when the wave peak reaches the end of the chain.
}
 \label{fig:FFT}
\end{figure}

We perform Fast Fourier Transform (FFT) on the velocity data to investigate the frequency components in the propagating wave in the graded HEC chain, as shown in Fig.~\ref{fig:FFT}. The red curves represent cutoff frequencies of the band structures, also denoted in Fig.~\ref{fig:FFT}(c) and (d) for the thinning and thickening chains, respectively. From Fig.~\ref{fig:FFT}(a) and (b), we observe that the power spectral density tends to follow the first and second passing bands of the dispersion curves (compare Fig.~\ref{fig:FFT}(a) and (c) for thinning and Fig.~\ref{fig:FFT}(b) and (d) for thickening HEC chain). This means that the propagating wave is mostly composed of the first and second vibrational modes of the HEC. The higher modes are not as evident as the lower ones, especially in the thickening chain, partly due to material damping of PLA.

Figure~\ref{fig:FFT}(e) and (f) show the spatial spectrum in the thinning and the thickening chain, respectively.
The spatial spectrum in the thinning HEC chain gradually includes high wavenumbers as time passes, which is consistent with the stiffening wave front as shown in Fig.~\ref{fig:vel_surf}(e). The widely expanded wave front with barely moving wave peak in the thickening HEC chain in Fig.~\ref{fig:vel_surf}(f) is well represented in Fig.~\ref{fig:FFT}(f). The wavenumber is more localized to low region as the wave width expands while the wide spectrum still exists even when the wave reaches the end for the localized peak.

\clearpage
\nocite{*}   
\bibliography{reference}

\end{document}